# Sub-40 nm resolution deep tissue imaging by image scanning emission saturation nanoscopy

Chenyi Wang[1], Tiange Zhang[1*],Chaohao Chen[2], Xuchen Shan[1], Meiqi Li[3], Xiaolan Zhong[1*], Fan Wang[1*]

[1] School of Physics, Beihang University, Beijing 100191, China

[2] School of Biomedical Engineering, Faculty of Engineering and Information Technology, The University of Technology Sydney, Sydney, NSW2007, Australia

[3] School of Life Sciences, Peking University, Beijing, China

*Corresponding author, E-mail: b24082@buaa.edu.cn, zhongxl@buaa.edu.cn, fanwang@ buaa.edu.cn

**Abstract**

The development of deep-tissue super-resolution imaging serves as an essential bridge toward non-invasive in vivo optical observation. However, there remain challenges to balance spatial resolution, imaging depth and phototoxicity. Here, we present a nanoscopy namely Image Scanning Emission Saturation (ISES) nanoscopy, achieving a lateral resolution of 37 nm, 1/25th of the excitation wavelength, at an imaging depth of 200 μm. Using a 976-nm doughnut-shaped excitation beam within an imaging-scanning microscopy configuration, we apply saturation-based point spread function (PSF) engineering and pixel-level confocal-pinhole enhancement to improve spatial resolution. As the high- and low-frequency components of the image OTF are concurrently acquired in a single scan via different camera pixels, Fourier-domain fusion can be employed with a single scanning dataset to further improve image quality. Compared with the traditional doughnut excitation beam-based adaptive pixel reassignment method, our strategy preserves the original frequency distributions and mitigates reconstruction artifacts in complex biological sample imaging. This strategy is generalizable and compatible with a variety of probes displaying saturation behavior. Beyond enabling a versatile and practical approach for deep tissue super-resolution imaging, it also informs the development of next-generation nanoprobes for imaging.

## 1. Introduction

By enabling high spatial-temporal resolution three-dimensional imaging, super-resolution microscopy has significantly revolutionized subcellular biology[1, 2], genomics[3, 4] and neuroscience[5, 6]. Most traditional super-resolution techniques investigate *in vitro* adherent cells, remaining a formidable challenge when imaging three-dimensional complex structures or biophysical interactions buried in deep tissues. The inevitable absorption and scattering originating from thick samples decrease both excitation and fluorescence intensities, impeding the application of Stimulated emission depletion microscopy (STED)[7, 8] and Single-molecule localization microscopy (SMLM) [9-11] for the drastic decrease of resolution with penetration depth. Other super-resolution technique, such as Structured Illumination Microscopy (SIM), is a cell-friendly method that provides a twofold resolution improvement.[12] However, its stripe patterns are disrupted by light scattering, absorption, and phase distortion, resulting in the artifacts of deep imaging.

To address the challenges of deep imaging, researchers have developed advanced imaging modalities that reduce excitation and emission photon requirements, correct optical aberrations, and employ nondiffracting beam shaping. MINFLUX and derived techniques[13, 14] mechanistically overcome the photon budget limitations inherent to single-molecule tracking–based nanoscopy. In parallel, post-processing-based denoising and deconvolution approaches, such as deepSTORM,[15] Deep-MSIM[16] and ROSE-3D[17], can efficiently reduce the number of frames for image reconstruction. Alternatively, integrating adaptive optics (AO) with nanoscopy can physically enhance both excitation and emission efficiencies by correcting optical aberrations, thereby improving spatial resolution and imaging penetration depth, though increasing the complexity of optical systems.[18, 19] Besides, the non-diffraction beam, including Bessel beam,[20] Airy beam and Lattice light-sheet[21, 22], has been used to overcome the scattering problem in deep-tissue due to its self-reconstructing characteristics. These efforts have pushed the achievable spatial resolution to 130 nm at imaging depths of up to 100 μm.[16] More recently, cost-effective nanoscopy techniques have been extended to deep-tissue imaging. Combining two-photon SIM with the line-scanning detection mode of the camera helps to detect finite structures at a penetration depth of 70 μm, with a resolution of 146 nm.[23] Image scanning microscopy (ISM), comparable with standard confocal systems, provides a good balance with the spatial resolution, image contrast, fidelity and depth. [24-28] By combining with the multi-focal[29] and spinning disk[30] strategy, the imaging depth of ISM has been enhanced up to 180 μm with a resolution of 144 nm.

Combining the emerging near-infrared imaging probe, called lanthanide-doped upconversion nanoparticles (UCNPs),[31] with the cost-effective nanoscopy, the resolution could be further improved. Owing to their multiphoton energy-transfer mechanisms, UCNPs display distinctive nonlinear emission saturation behavior. A single-beam excitation super-resolution strategy named near-infrared emission saturation nanoscopy (NIRES) [32]achieved a 50-nm lateral resolution when imaging a single UCNP through 93 μm of liver tissue under relatively low excitation intensity. Nevertheless, further improvement of the resolution meets challenges due to the emission saturation effect or the optical system. Moreover, single doughnut-beam scanning microscopy exhibits a missing frequency component in its optical transfer function (OTF), ultimately restricting its effective resolution and causing deconvolution artifacts when imaging continuous samples. [33]Alternatively, the Gaussian-like and doughnut-shaped PSF can be acquired by two separate scans[34] or different saturated emissions associated with different energy levels, [35] inevitably leading to an increase in system complexity. Hence, more investigation is needed to achieve sub-50-nm resolution in deep-tissue imaging using a simplified optical setup.

In this work, we present a new deep tissue super-resolution imaging strategy, namely image scanning emission saturation microscopy (ISES), which uses a confocal system with a phase mask and a detection camera (Figure 1b-iii). In the experiments, the 800-nm emission was selected to reduce scattering in deep tissue, which becomes saturated at a certain excitation intensity, tuning the inner full-width at half -maximum (iFWHM) of the doughnut-shaped excitation beam to be smaller. Taking advantage of the near-infrared excitation and emission as well as the mechanism of the ISM, we further enhanced the resolution and the imaging depth. To demonstrate super-resolution of ISES in the frequency domain, we utilized PSFs of different camera pixels in the ISES image stacks and successfully reconstructed Gaussian-like and doughnut-shaped OTFs in a single scan. In this way, we successfully reconstructed high-fidelity super-resolution images of the cytoskeletons in HeLa cells. Our work provides a simple one-scan and universal method to engineer a high resolution in both the spatial and frequency domains, as well as optimize spatial information in cellular structures by using only a single near-infrared laser.

## 2. Results and discussions

### 2.1 The principle of ISES

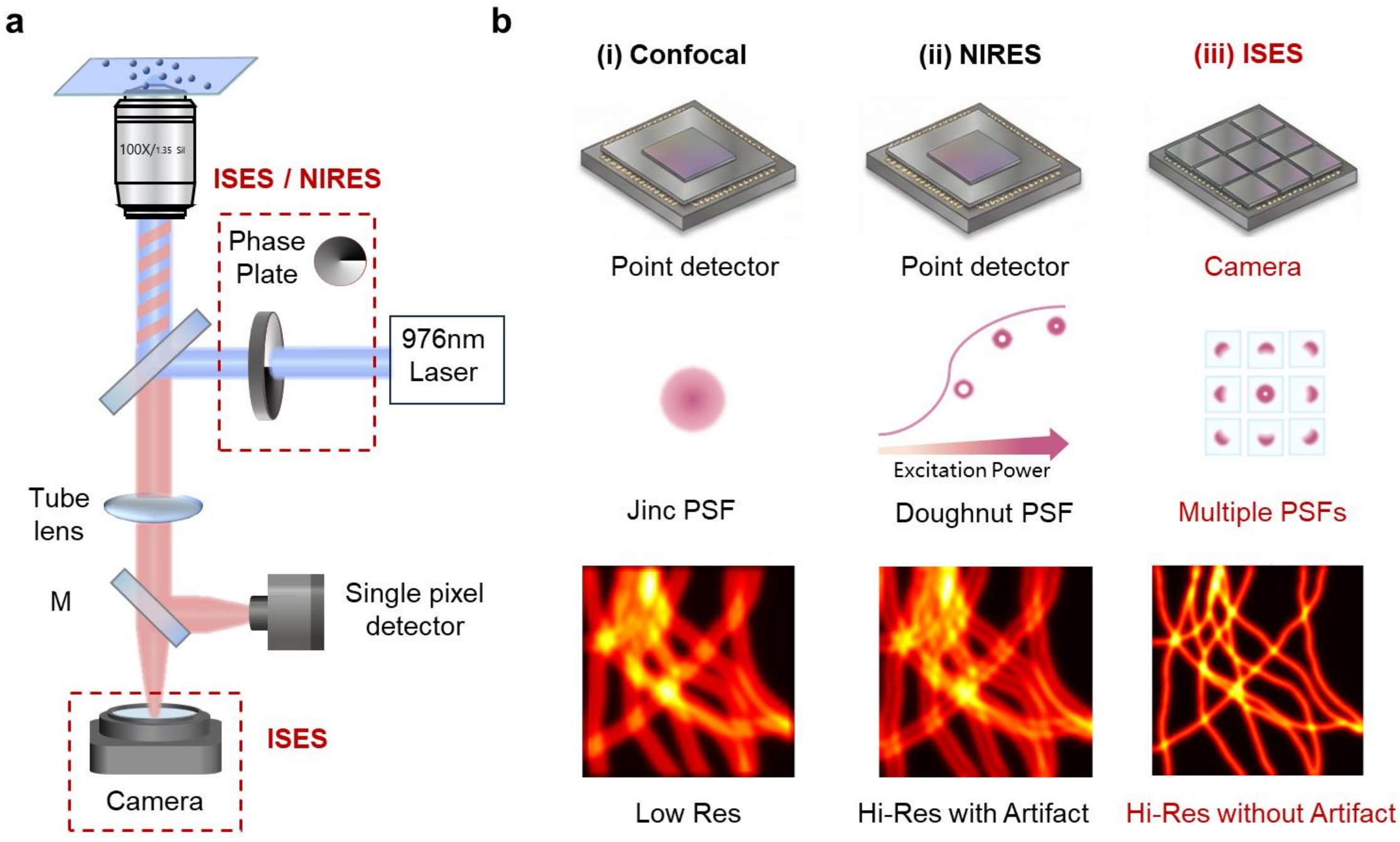

**Figure 1 | Principle of Imaging Scanning Emission Saturation nanoscopy (ISES).** a, Optical setup of ISES. A Vortex Phase Plate (VPP) was placed to modulate the 976-nm excitation beam to the doughnut shape. Then, the doughnut beam passed through the dichroic mirror (DM) and was focused by the objective lens on the sample plane. The fluorescence from the sample is collected and imaged on the SPAD array camera via the tubelens and a pair of relay lenses, which are not shown here. b, Schematic of confocal, NIRES and ISES nanoscopy. Traditional confocal and NIRES microscopy both utilize a point detector. The fluorophore of NIRES nanoscopy shows a unique nonlinear response to the excitation power. With the increase of the excitation power, the slope of the excitation-emission dependent curve gets lower than 1. Our proposed ISES nanoscopy further replaces the point detector with a SPAD array camera. As the emission saturation phenomenon prompts a narrower iFWHM of the detected PSF, the SPAD array camera further enhances the resolution of the detected image by utilizing smaller pinholes and the Fourier fusion algorithm.

The ISES system can be readily implemented by modifying a conventional confocal microscope (Fig. 1a), in which a vortex phase plate (VPP) is introduced to generate a doughnut-shaped excitation point-spread function at the focal plane. Unlike confocal and NIRES microscopy, ISES requires a camera-based detector to realize an image-scanning microscopy detection scheme, where each pixel of the camera is considered as an individual pinhole to collect the same sample region from a different perspective after a complete scan (Figure 1b-iii). Compared with the traditional diffraction-limited confocal microscope, the doughnut excitation PSF of NIRES microscopy (Figure 1b-ii) can be modulated by the emission saturation of the fluorophores. When increasing the excitation intensity, the dip of the detected PSF gradually narrowed, enabling super-resolution microscopy. Nevertheless, the doughnut-based modulation of NIRES induced an absence in certain frequency components, resulting in a decrease in resolution when imaging the continuous samples. In the proposed ISES strategy, imaging resolution is enhanced while reconstruction artifacts are simultaneously suppressed. The raw ISES image stack forms a multidimensional dataset, in which each scanned image corresponds to an individual detector pixel. By treating these pixels (with an effective size of 0.2 AU) as confocal pinholes, the system achieves improved spatial resolution. The adaptive pixel reassignment (APR) algorithm[36] applied to the camera pixels further enables concurrent optimization of resolution and signal-to-noise ratio. Importantly, pixels displaced from the optical axis capture distinct spatial frequency components, thereby encoding complementary optical transfer function (OTF) information within a single scan. Merging these data in the Fourier domain compensates for the missing frequency components associated with a single doughnut-shaped OTF, thereby mitigating information loss and improving the fidelity of the reconstructed image.

## 2.2 The simulation of high-frequency enhancement on ISES

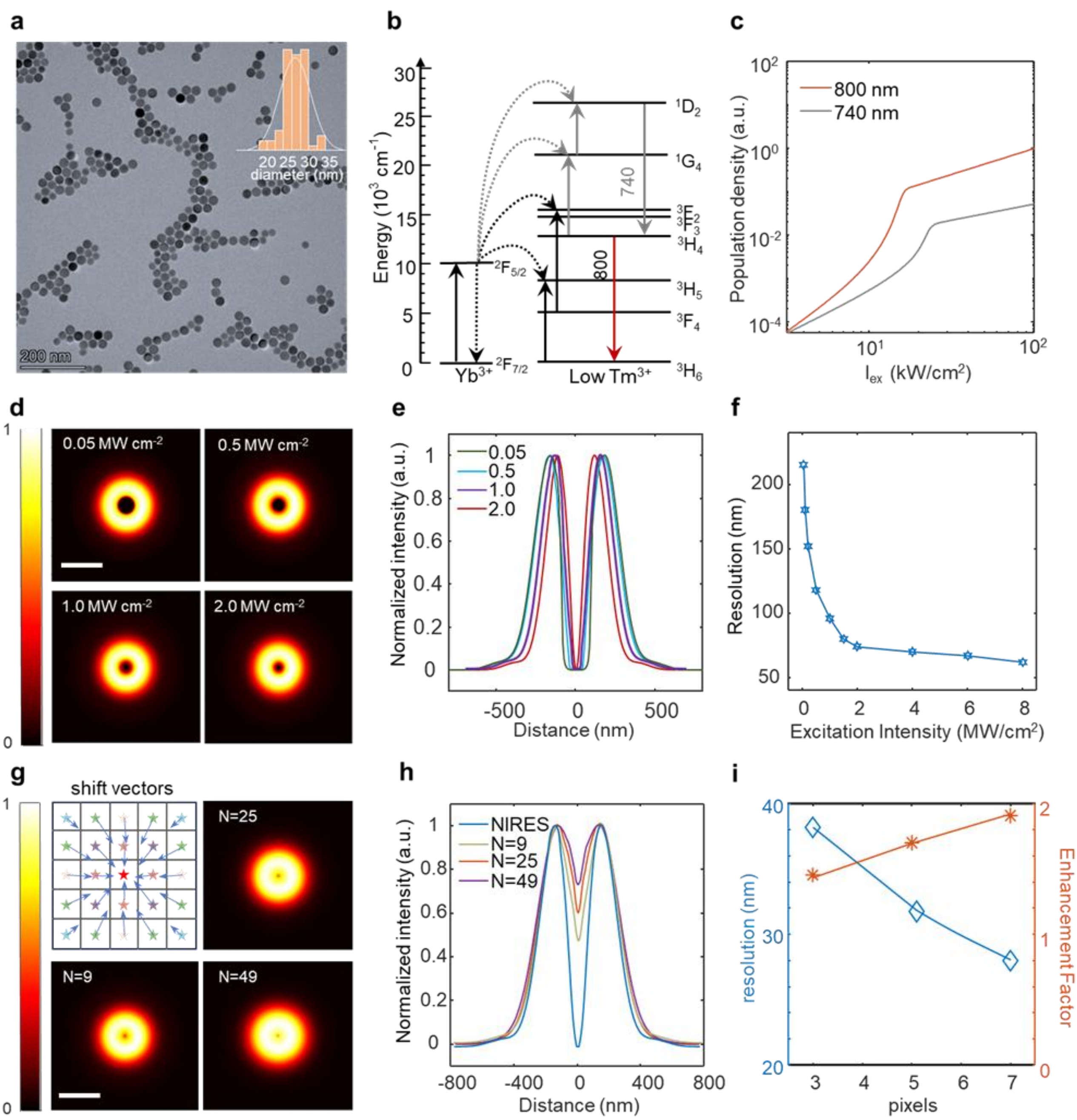

**Figure 2 | Simulations for ISES in $Yb^{3+}/Tm^{3+}$ co-doped nanoparticles.** a, Energy diagram of the Yb-Tm rate equation model. b, The simulated intensity curves of 800-nm and 740-nm emissions versus 976-nm excitation power density in Yb-Tm nanoparticles. c, Transmission electron microscopy image of the $NaYF_4$: Yb/Tm(20/8%) @$NaYF_4$ nanoparticles, with an average size of $\approx$28.8 nm. d-e, The simulated "negative" contrast emission PSFs of a single UCNP and their cross-section profiles at four different excitation powers of 0.05, 0.5, 1.0, and 2.0 MW cm$^{-2}$, respectively. Theoretically, the obtained detected PSF of the single UCNP is regarded as the convolution of the effective PSF and the size of the UCNP. Pixel size is 10 nm. Scale bar is 500 nm. e, Cross-section profiles of the saturated upconversion emission of UCNPs at four different excitation powers of 0.05, 0.5, 1.0, and 2.0 MW cm$^{-2}$. f, The resolution of UCNPs as a function of the excitation intensity. g-h, The simulated "negative" contrast emission PSF and the cross-section profiles of a single UCNP, respectively, reconstructed with different numbers of pixel elements (N). The sub-images are shifted to the center position with different shift vectors by the traditional APR algorithm. Pixel size is 10 nm. Scale bar is 500 nm. i, The resolution of ISES and its enhancement factor to the resolution of NIRES as a function of the number of pixels (3 ×3, 5×5, 7×7). The excitation intensity is 2.0 MW cm$^{-2}$.

To investigate the ISES strategy, we first performed theoretical modeling analysis. Hexagonal-phase $NaYF_4$:Yb/Tm (20/8%) @$NaYF_4$ nanoparticles with an average diameter of 28.8 nm were synthesized as the imaging probes (TEM images shown in Fig. 2a) and the modeling target, as this doping concentration yields strong near-infrared emission. Through the multi-photon energy transfer pathways between $Yb^{3+}$ and $Tm^{3+}$, various emission bands ranging from the near-infrared to visible region are exhibited. The near-infrared excitation and emission ensure the penetration depth in deep tissues. The detailed description of the energy transfer between $Yb^{3+}$ and $Tm^{3+}$ is given using rate equations (Figure 2b and Supplementary Figure 3). After characterizing the curves of emission intensity versus excitation intensity, clear saturation thresholds of 800-nm emission appeared at about 0.2 MW $cm^{-2}$, exhibiting an unusual dependence on pumping intensity (slope $<1$). Further increasing the excitation intensity, the emission intensity tends to increase slowly (Figure 2b). The saturation process was attributed to the limited sensitizer and emitter numbers. Among the two main near-infrared emission bands, the 800 nm band exhibits a lower saturation threshold than the 740 nm band (Figure 2c). This behavior arises because the 740 nm emission originates from the four-photon-excited $^1D_2$ level, which requires additional transitions before saturation, whereas the 800 nm emission is associated with the two-photon-excited $^3H_4$ level. Since lower saturation threshold results in higher resolution for NIRES, the 800-nm emission was selected for further imaging. A tightly-focused doughnut-shaped excitation beam is first generated as the excitation PSF (Supplementary Figure 4). According to the definition of resolution in NIRES microscopy,[32] its effective PSF can be calculated through the multiplication between the excitation PSF and the 800-nm emission modulated PSF. As the excitation intensity aggrandized, the dip narrowed, and a 56-nm lateral resolution was obtained at the excitation intensity of 2 MW $cm^{-2}$ through a point detector (Figure 2d,e). Further increasing the excitation intensity, the UCNPs become oversaturated, restricting the improvement of resolution (Figure 2f and Supplementary Figure 4).

To further explore our ISES strategy, we then replace the point detector with a camera equipped with different numbers of virtual pixel elements. After a complete scan, each element acts as a smaller confocal pinhole and individually builds a detected PSF, contributing to collecting more high-frequency components (Figure 2g and Supplementary Figure 5). The PSF of the central element maintains the doughnut shape but is comparatively dim, whereas other misaligned elements are either asymmetric or completely lose the doughnut shape. By setting the number of pixel elements to N=9, a 38-nm lateral resolution is obtained with an enhancement factor of 1.47 in contrast with NIRES microscopy (Figure 2g, h). According to the cross-section profile (Figure 2h), the increase in the number of pixel elements induces higher resolution. However, merging a large number of pixel elements results in an increased dip value in the PSF, thereby degrading effective resolution. This increase originates from the incorporation of low spatial frequency components from off-center pixels during the APR process. Notably, the gain in effective resolution becomes marginal when the dip value exceeds 0.5. Therefore, N=9 was selected for subsequent experiments.

## 2.3 Characterization of ISES reveals sub-40-nm resolution

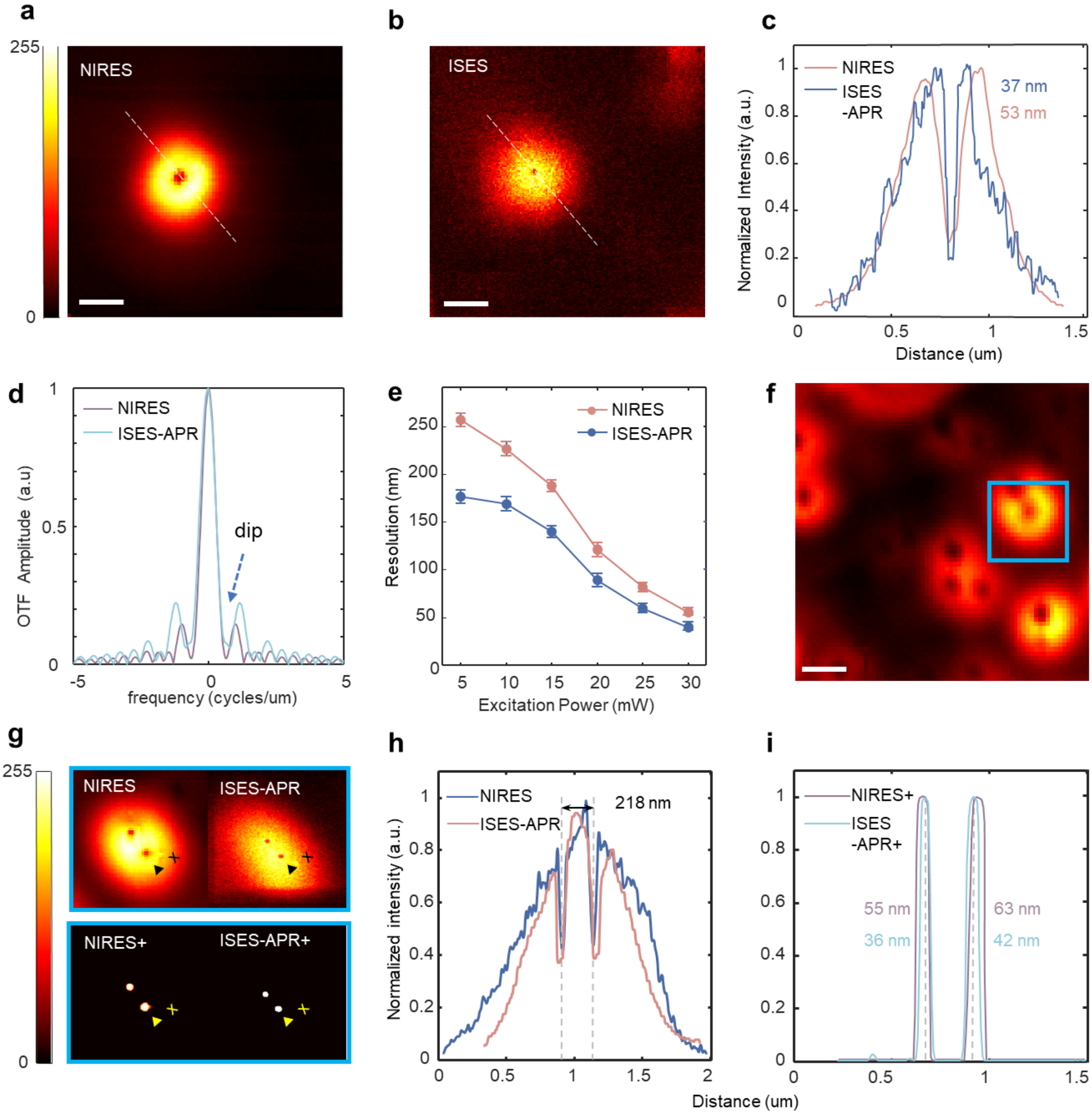

**Figure 3 | ISES for super-resolution imaging.** a-b, Comparison of NIRES and ISES image of a single UCNP at an excitation intensity of 1.5 MW cm$^{-2}$. The ISES image is reconstructed by using the central 3×3 pixels of the scanned raw images. Pixel dwell time: 5 ms for NIRES and 20ms for ISES; pixel size: 25 nm for NIRES and 10 nm for ISES. Scale bar is 500 nm. The comprehensive overview of the post-processing of ISES consists of the following steps: (1) subtracting the dark count map, (2) reorganizing and extracting the sub-images, (3) positioning the pixel-level center, (4) applying pixel reassignment. c-d, Normalized intensity and OTF amplitude cross-section profiles of images a-b along the white dash lines, respectively. e, The average NIRES and ISES resolutions of 5 nanoparticles as a function of the excitation intensity. Error bars are defined as s.d. from the line profiles of several measurements. f, NIRES images of a 6 µm × 6 µm area containing clusters of UCNPs at an excitation intensity of 0.2 MW cm$^{-2}$. Pixel Size: 30 nm, Pixel dwell time: 5ms. Scale bar is 1µm. g, The zoom-in NIRES image of an area of interest marked by the blue square from f, the corresponding ISES image and the processed data of NIRES+ and ISES+ image, at an excitation intensity of 1.8 MW cm$^{-2}$. h-i, Normalized intensity cross-section profiles of the “negative” and “positive” images in g along the lines, respectively.

We further validate the ISES strategy through experimental imaging of single $NaYF_4$:Yb/Tm (20/8%)@$NaYF_4$ nanoparticles. The curve of emission intensity versus excitation intensity was fitted well with the simulated result, revealing the emission saturation effect with an S-shaped curve (Supplementary Figure 6). To reveal the diffraction-limited resolution of the system, we measured the confocal image of the 800-nm emission with an effective diameter of 0.55 Airy Units (AU), which has an FWHM of 488 nm (Supplementary Figure 7). Subsequently, we implemented the effective PSFs of single nanoparticles as a function of excitation intensity based on the statistics of five selected single nanoparticles. Due to this optical nonlinear response, the detected PSF shrinks drastically, with lateral iFWHMs of 53 nm at the excitation intensity of 1.5 MW $cm^{-2}$ (Figure 3a,c). To further improve the resolution, we acquired ISES image stacks based on the APR algorithm. By switching the detector to the SPAD array camera with 32×32 pixels, the pixel containing the beat symmetry of the doughnut shape PSF was positioned as the "central" pixel. Subsequently, its nearby 3×3 pixels were utilized for pixel reassignment for compensation of resolution enhancement and a reasonable SNR. A 37-nm lateral resolution $(\lambda/20)$ can be reached, with an enhancement factor of 1.43 (Figure 3b,c). Compared with NIRES, the high-frequency component of OTF exhibits a remarkable increase (Figure 3d). The increased excitation intensity results in better resolution, and the ISES-APR nanoscopy always exhibits superior resolution (Figure 3e). Furthermore, we examined the ability of ISES-APR for resolving a single nanoparticle from UCNPs clusters. In a typical area of the NIRES image marked by a blue square, two adjacent nanoparticles within a diffraction limit area were indistinguishable under low excitation intensity (0.2 MW $cm^{-2}$, Figure 3f). In relatively high excitation intensity (1.8 MW $cm^{-2}$), both NIRES and ISES-APR can resolve a single nanoparticle by either negative or positive images (Figure 3g), but the positive image (ISES-APR+) provides the optimal resolving quality for clearly illustrating the sub-40 nm resolution with a distance of 218 nm (Figure 3h, i). While spatial resolution is enhanced, the OTF retains a low-frequency dip (Figure 3d), which suppresses low-spatial-frequency information and reduces the effective resolution for complex samples. This drawback can be solved by incorporating the Fourier-domain fusion as shown below.

## 2.4 One-scan Fourier fusion frequency-domain imaging concept

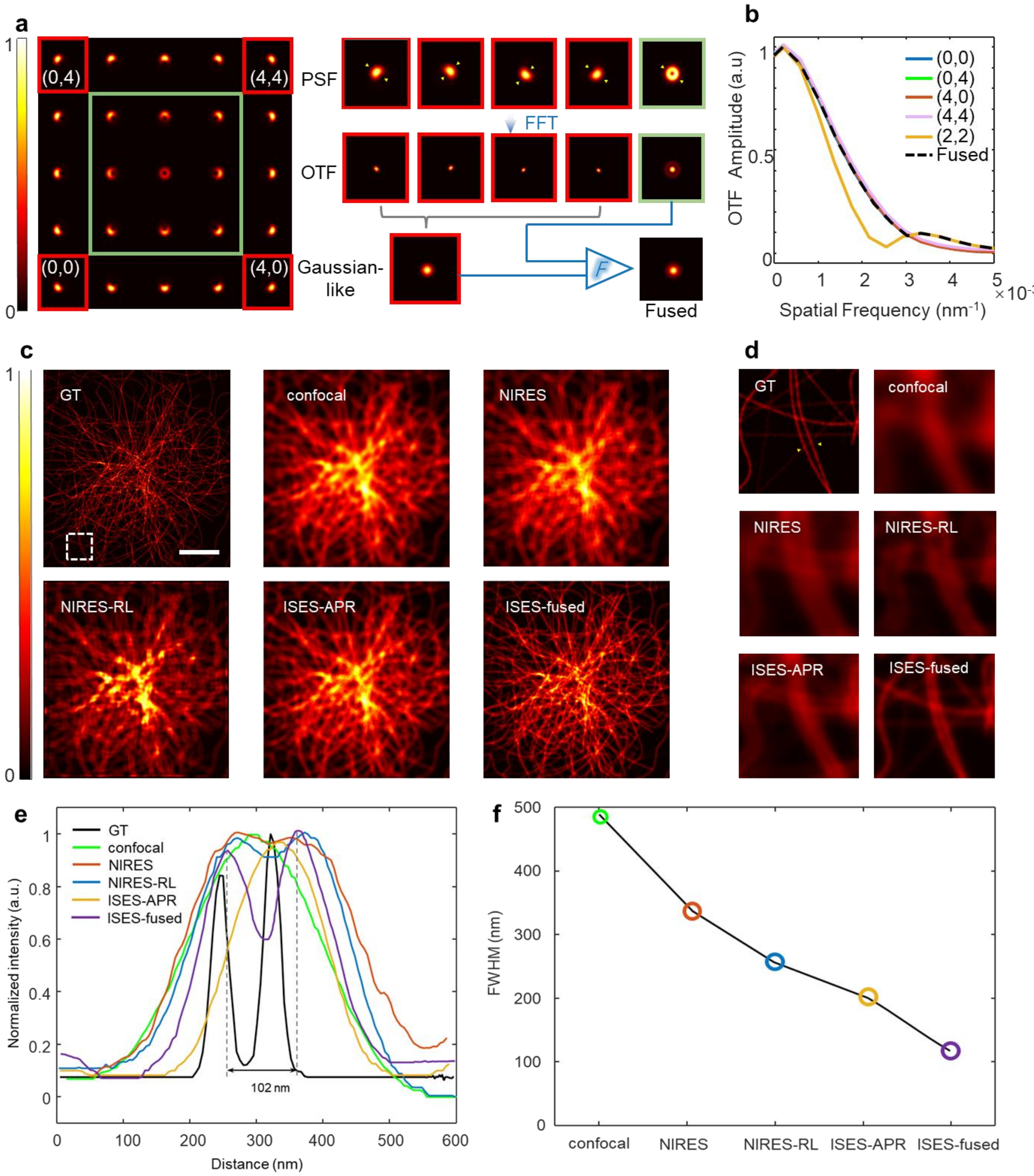

**Figure 4 | The enhanced image quality utilizing the one-scan Fourier fusion method in ISES.** a, The principle of the one-scan Fourier fusion method. The OTFs of the PSFs in 5×5 pixels can be acquired after fast Fourier transform. The doughnut-shaped OTF appears in the central pixel, and the Gaussian-like OTF is generated after summing the OTFs of the four corner pixels ((0,0),(0,4),(4,0),(4,4)). Subsequently, by selecting the cut-off frequency, a fuse OTF is generated by selecting the low frequency component of the Gaussian-like OTF and the high frequency component of the doughnut OTF. Note that the zero spatial frequency starts from the center of each image. b, Normalized OTF amplitude cross-section profiles of sub-images in four corner pixels, the central pixel and the fused image in a, respectively. c, The ground truth of the microtubule image and the corresponding confocal, NIRES, NIRES-RL, ISES-APR, and ISES-fused images. Scale bar is 2 μm. d, Magnified views of sub-regions in the ground truth, confocal, NIRES, NIRES-RL, ISES-APR, ISES-fused images in c. e, Intensity profiles along the lines in d. f, The effective resolution comparison for the five methods in e, which is acquired by Gaussian fitting.

Enhancement of the effective resolution in ISES relies on Fourier-domain fusion of low-spatial-frequency components, which is conventionally acquired via an additional Gaussian-beam scanning process.[34]In this work, we demonstrate that the same functionality can be achieved using only a single scan. Note that the number of pixel elements was set to be N=25 (5×5) and each pixel generates different PSFs during scanning. Due to the mismatch between excitation and detection, the OTFs obtained from the four corner pixel elements ((0,0), (0,4), (4,0), and (4,4)) contain substantial low-spatial-frequency information, and their summation produces a Gaussian-like OTF (Figure 4a and Supplementary Figure 8, 9). In contrast, the central pixel element (2,2) and the PSF generated by the APR algorithm remain doughnut-shaped. As a result, both low- and high-spatial-frequency components can be acquired in a single scan, enabling Fourier-domain fusion. The normalized line profiles of the OTFs in the above five channels are shown in Figure 4b. The 1D frequency responses of the four Gaussian-like channels are almost the same in four different directions marked in Figure 4a. However, the central doughnut OTF still lacks certain components that are contained in the passband of the Gaussian-like channels. The cut-off frequency ($f_{cut}$) is set at the cross-point between the Gaussian and doughnut OTF line profile, acting as a low-pass and a high-pass binary mask to Gaussian and doughnut channels, respectively. (Figure 4b). More details of the fusion process are described in the Supporting Information. A simulation dataset of synthetic microtubules was used as ground truth to validate the impact of the Fourier fusion method on super-resolution reconstruction of the samples with dense structures (Figure 4c). The magnified sub-region, marked by a white square, is analyzed at the positions indicated by a line, which contains two adjacent microtubules (Figure 4d). For the confocal scanning image, the two microtubules are blurred due to the diffraction-limited resolution. The details by NIRES are hindered by artifacts. After Richard-Lucy deconvolution, the artifact still exists, leading to the mistake in the positions of the microtubules (Figure 4c,d). After reconstruction through the APR algorithm, despite the improvement in clarity, the detailed structures remain overlapped and the two adjacent microtubules cannot be distinguished. In contrast, the Fourier fusion process achieves the subtlest details without major positioning mistakes. Two adjacent microtubules with a 108-nm distance marked by white triangles in GT can be resolved only by the fusion technique (Figure 4e). With the smallest FWHM compared with deconvolution or APR algorithm (Figure 4f), our findings provide direction for one-scan frequency domain super-resolution bioimaging to reconstruct high-resolution and high-fidelity images.

## 2.5 Super-resolution imaging under deep-tissue and cells

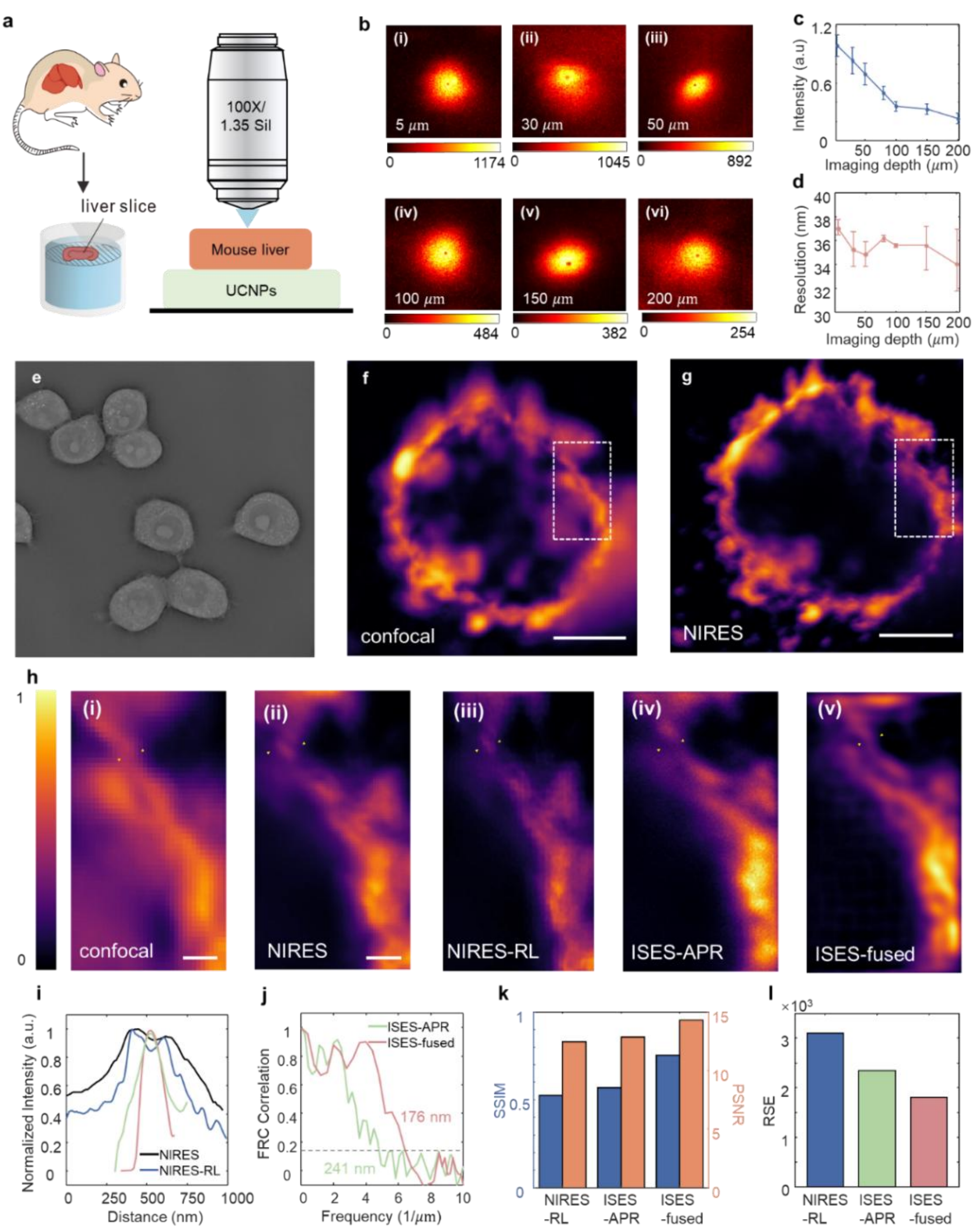

**Figure 5 | High-resolution ISES imaging in different imaging depths and subcellular filaments.** a, Illustration of the mouse liver slices covering onto the UCNPs on the slides. b, Single particle imaging of ISES at different depths in liver tissue. Pixel size is 10 nm. c, The normalized emission attenuation at different depths through the liver tissue. d, The corresponding FWHM in b; the resolutions of ISES in d are 37.2 ± 1.3 nm (5 μm), 35.1 ± 2.2 nm (30 μm), 34.6 ± 1.9 nm (50 μm), 35.8 ± 0.4 nm (100 μm), 35.4 ± 2.5 nm (150 μm), 34.1 ± 2.8 nm (200 μm). e, Bright field imaging of fixed HeLa cells immunolabeled with phalloidin-conjugated UCNPs. f-g, Confocal and NIRES images of a 28 μm × 28 μm area containing the labeled f-actin of a HeLa cell. The scale bar is 6 μm. Pixel size is 200 nm in f and 70 nm in g, respectively. h, (i) and (ii) are the zoom-in 5 μm × 12 μm areas of interest from f and g, respectively. The scale bar is 800 nm. (iii) is the processed data of (ii) using the Richardson-Lucy deconvolution algorithm. (iv) and (v) are the processed ISES images of the same area using traditional APR and the Fourier fusion algorithm, respectively. i-j, Intensity profiles and resolution comparison obtained by FRC analysis of NIRES, NIRES-RL, ISES-APR, ISES-fusion along the directions in h, respectively. k-i, Column chart of PSNR, SSIM and RSE comparison between NIRES-RL, ISES-APR and ISES-fused.

We further demonstrate the capability of ISES in deep-tissue imaging and in resolving subcellular features. The UCNPs is embedded beneath mouse liver tissue slices (Figure 5a). By slightly increasing the excitation intensity to compensate for the light absorption of the excitation beam, both the positive and negative contrast images can maintain sub-40 nm resolution through the liver slices with a thickness from 5 μm to 200 μm (Figure 5b-d and Supplementary Figure 11,12). Due to the lower refractive index and scattering for NIR emission, 21.6% detectable signal of 800-nm emission remains through a 200-μm tissue slice (Figure 5c). Subsequently, we performed subcellular imaging of the cytoskeletons. The UCNPs get surface modified with polyacrylic acid (PAA) to render aqueous dispersity and then conjugated with phalloidin (Supplementary Information). The actin filaments of fixed HeLa cells were immunolabeled with phalloidin-conjugated nanoparticles (Figure 5e, f). NIRES imaging (Figure 5g) of the actin filaments in a whole cell with a field of view 28 μm × 28 μm shows an obviously improved resolution compared with the confocal imaging (Figure 5f). Nevertheless, the artifacts are induced by the frequency loss of the doughnut-beam scanning of NIRES. Following Richardson–Lucy (RL) deconvolution applied to a small field of view (5 μm × 12 μm; marked by the white square in Figure 5e), the resolution of individual filaments can be enhanced, but artifacts persist. Notably, checkerboard artifacts become more severe (Figure 5h(iii)). In parallel, we performed the ISES image stacks of the same area and separately processed them with the conventional APR algorithm and the proposed Fourier fusion method (Figure 5 h (iv), (v)). Both of them were reconstructed utilizing the central 5×5 sub-images. Compared with the RL deconvolution and APR algorithm, the Fourier fusion method preserves most of the structural information, enabling a drastically improved effective resolution of 176 nm evaluated by Fourier Ring correlation (FRC) (Figure 5i, j). Furthermore, image quality was assessed with NanoJ-SQUIRREL. [37] The Fourier fusion method achieved the highest peak signal-to-noise ratio (PSNR) and structural similarity (SSIM) as well as the lowest resolution scaled error (RSE) in quantitative comparisons (Figure 5k, l), indicating the best performance in suppressing the artifacts and realizing high-fidelity super-resolution reconstruction.

## 3. Conclusion

By utilizing the emission saturation effect of $Yb^{3+}/Tm^{3+}$ co-doped nanoparticles, we report ISES super-resolution approach. Offering the advantage of simplicity and stability, ISES achieves sub-40 nm spatial resolution. Each pixel element in our SPAD array camera works individually and measures only 0.2 A.U., which contributes to enhanced resolution. Benefiting from the low absorption and scattering of the near-infrared excitation and emission, our ISES approach realizes an imaging depth of 200 μm through mouse liver slices. By applying Fourier fusion techniques, ISES acquires Gaussian-like and doughnut-shaped PSF patterns through only a single scan, which overcomes the drawback of doughnut-beam excitation for the absence of certain frequency components in its passband, enabling high-resolution and high-fidelity reconstruction of actin filaments in HeLa cells, underscoring its potential for visualizing complex structures in deep tissues and cells. In summary, ISES is a robust approach to enable super-resolution in both the spatial and frequency domains, which offers new insights into biological exploration in subcellular structures and complex dynamics in deep tissues and delivers new pathways into super-resolution *in-vitro* and *in-vivo* imaging. Replacing the objective with a longer working distance or shifting the emission wavelength to the second near-infrared region may further improve the imaging depth. Optimizing the component of the UCNPs may decrease the emission saturation threshold, which further improves the emission intensity and the spatial resolution. Notably, ISES nanoscopy is a robust method that is comparable with other techniques meaning that the resolution and penetration depth can be further improved. Integrating adaptive optics to correct the sample-induced aberrations may allow higher-fidelity imaging in more complex biological samples like organs and small animals.[38] In parallel, the development of machine-learning-based algorithms may reduce the requirement of photon budget or shorten the scanning and reconstruction time.[39] The use of near-infrared-enhanced cameras[40] can further improve photon collection efficiency, particularly in the near-infrared region, leading to enhanced image quality. A potential limitation of ISES is its suboptimal temporal resolution, arising from the relatively long lifetimes of UCNPs. However, as the strategy is compatible with a broad range of fluorophores exhibiting emission saturation, this constraint could be alleviated through appropriate selection and optimization of emitters[41, 42].

## Methods

Optical spectroscopic and microscopic setup, synthesis and characterizations of lanthanide-doped nanoparticles, optical measurement and reconstruction details, theoretical modeling and simulations and any other associated figures and references are provided in the Supplementary Information.

## Data availability

The source datasets generated and analyzed during the current study are available from the corresponding authors upon reasonable request.

## Code availability

The codes for theoretical modeling and reconstruction of ISES are available from the corresponding authors upon reasonable request.

**Acknowledgments:** This work was supported by the Beijing Natural Science Foundation(1232027) and National Natural Science Foundation of China (U23A20481,62275010, and 62405015). We are grateful to the Atomic-Scale In Situ Fabrication platform of the Analysis & Testing Center of Beihang University for the facilities, and the scientific and technical assistance.

**Competing interests:** The authors declare no competing interests.